\documentclass[aps,prd,twocolumn,longbibliography,showkeys,showpacs,preprintnumbers,superscriptaddress,nofootinbib,10pt]{revtex4-1}
\usepackage{graphicx,epsf,bm,amsmath,amsfonts,amssymb,epstopdf,natbib,hyperref,color,verbatim,multirow,bm,mathrsfs,appendix,subcaption}
\usepackage{float}
\usepackage[justification=raggedright,singlelinecheck=false]{caption}
\hypersetup{colorlinks=true,urlcolor=blue,citecolor=blue,linkcolor=blue,menucolor=blue,anchorcolor=blue,filecolor=blue}
\date{\today}

\begin{document}

\title{Machine Learning-Based Analytical Expressions for Gray-Body Factors and Application to Primordial Black Holes}

\author{Guan-Wen Yuan}
\email{guanwen.yuan@unitn.it}
\affiliation{Department of Physics, University of Trento, Via Sommarive 14, 38123 Povo (TN), Italy}
\thanks{G.Y and M.C. contributed equally to this work}
\affiliation{Trento Institute for Fundamental Physics and Applications (TIFPA)-INFN, Via Sommarive 14, 38123 Povo (TN), Italy}

\author{Marco Calz\`{a}}
\email{marco.calza@unitn.it}
\affiliation{Department of Physics, University of Trento, Via Sommarive 14, 38123 Povo (TN), Italy}
\thanks{G.Y. and M.C. contributed equally to this work}
\affiliation{Trento Institute for Fundamental Physics and Applications (TIFPA)-INFN, Via Sommarive 14, 38123 Povo (TN), Italy}

\author{Davide Pedrotti}
\email{davide.pedrotti-1@unitn.it}
\affiliation{Department of Physics, University of Trento, Via Sommarive 14, 38123 Povo (TN), Italy}
\affiliation{Trento Institute for Fundamental Physics and Applications (TIFPA)-INFN, Via Sommarive 14, 38123 Povo (TN), Italy}
\affiliation{School of Mathematics and Statistics, University of Sheffield, Hounsfield Road, Sheffield S3 7RH, United Kingdom \looseness=-1}

\begin{abstract}

\noindent Symbolic Regression (SR) is a machine learning approach that explores the space of mathematical expressions to identify those that best fit a given dataset, balancing both accuracy and simplicity. We apply SR to the study of Gray-Body Factors (GBFs), which play a crucial role in the derivation of Hawking radiation and are recognized for their computational complexity. We explore simple analytical forms for the GBFs of the Schwarzschild Black Hole (BH). We compare the results obtained with different approaches and quantify their consistency with those obtained by solving the Teukolsky equation. As a case study, we apply our pipeline, which we call \texttt{ReGrayssion}, to the study of Primordial Black Holes (PBHs) as Dark Matter (DM) candidates, deriving constraints on the abundance from observations of diffuse extragalactic $\gamma$-ray background. These results highlight the possible role of SR in providing human-interpretable, approximate analytical GBF expressions, offering a new pathway for investigating PBH as a DM candidate.  

\end{abstract}

\maketitle

\section{Introduction}
\label{sec:introduction}
The scientific discovery process often begins with the identification of empirical patterns in the data. For example, Kepler's third law of planetary motion was formulated through careful analysis of decades of observational data collected by Tycho Brahe~\cite{hawking2002shoulders}. Similarly, Planck's radiation law, a cornerstone of quantum theory, did not originate from first principles, but emerged from fitting a symbolic form to experimental data~\cite{stewart2016blackbody}. Edwin Hubble's inference of the expanding universe relied likewise heavily on his intuition within noisy datasets~\cite{hubble1929relation}. Although these historical breakthroughs were achieved through guess work or intuition of remarkable human insight, discovering compact symbolic relationships remains a challenging and time-consuming task.  With the advent of large and high-dimensional datasets, uncovering concise analytical expressions may appear 
daunting without automated tools. This challenge motivates the use of Symbolic Regression (SR), a supervised machine learning approach that seeks analytic expressions to describe the data by minimizing both prediction error and model complexity. Unlike traditional machine learning methods that fit parameters within fixed model structures, SR explores a vast space of symbolic forms to identify analytic, human-interpretable models~\cite{camps2023discovering}.

By defining a target variable, a collection of potential explanatory variables, and a set of available mathematical operators, SR aims to find equations that accurately predict the target variable as a function of the explanatory ones, while remaining concise and interpretable~\cite{makke2024interpretable,NEURIPS2022_dbca58f3,wang2023scientific}. Various SR frameworks are currently accessible, such as \texttt{EUREQA}~\cite{schmidt2009distilling}, \texttt{Operon}~\cite{burlacu2020operon}, \texttt{DSR}~\cite{petersen2021deepsymbolicregressionrecovering}, \texttt{EQL}~\cite{sahoo2018learning}, \texttt{ESR}~\cite{Bartlett:2022kyi}, and \texttt{AI Feynman}~\cite{Udrescu:2019mnk}. These interpretable machine learning tools have recently sparked renewed interest, catalyzing notable progress across both theoretical and applied sciences~\cite{2021arXiv210714351L,Bernal:2021ylz,Butter:2021rvz,shao2022finding,li2023electron,Bartlett:2023gvh,angelis2023artificial,Koksbang:2023wez,Koksbang:2023guu,Desmond:2023wqf, Sousa:2023unz, Bartlett:2023cyr, Bartlett:2023gvh, Koksbang:2023sab, Thing:2024qja,Bartlett:2024jes, Constantin:2024dpq, Sui:2024wob,Makke:2025vmd,Morales-Alvarado:2024jrk,2024arXiv240203115S,Tsoi:2024pbn}. 
In this work, we use \texttt{PySR}~\cite{Cranmer:2020wew, cranmer2023interpretablemachinelearningscience}, an open source SR library that effectively combines genetic algorithms with gradient-based optimization to learn symbolic models. This package has been widely applied in both astronomy and cosmology to find analytical expressions hidden in data, such as modeling the distribution of neutral hydrogen~\cite{Wadekar:2020oov}, finding universal relations in subhalo properties~\cite{Shao:2021qoa}, searching the galaxy-halo connection~\cite{Delgado:2021cuw}, reducing the Sunyaev-Zeldovich flux-mass scatter~\cite{Wadekar:2022cyw}, exploring feedback from active galactic nuclei~\cite{Wadekar:2022bss}, discovering black hole (BH) mass scaling relations~\cite{2023arXiv231019406J}, rediscovering orbital mechanics~\cite{2023MLS&T...4d5002L}, investigating the dynamical dark energy~\cite{Sousa-Neto:2025gpj} and so on.

Here, we apply SR to the study of Gray-Body Factors (GBFs) in Primordial Black Holes (PBHs), which are compelling Dark Matter (DM) candidates~\cite{Carr:2020xqk,Bird:2022wvk,Green:2024bam}. Despite decades of experimental effort, the nature of DM remains unknown~\cite{Bertone:2004pz, Feng:2010gw, Cirelli:2024ssz}. Several well-motivated candidates have been proposed, such as weakly interacting massive particles~\cite{Roszkowski:2017nbc,Yuan:2021mzi}, axion/axion-like particles~\cite{Adams:2022pbo,Yuan:2020xui}, fuzzy dark matter and ultralight dark matter~\cite{Hu:2000ke,Yuan:2022nmu}. Among these possibilities, PBHs are unique since their origin does not require beyond-Standard Model physics: they can form naturally in the early Universe from the collapse of large density fluctuations~\cite{Hawking:1971ei, Carr:2016drx,Yuan:2023bvh}. 
PBHs are unique in that they may span an extremely broad mass range, $10^{-5} \sim 10^{55}$g~\cite{Carr:1974nx, Carr:1975qj}, distinguishing them from BHs formed through stellar evolution.
A key theoretical ingredient in modeling PBHs evaporation is the GBFs, which quantify the frequency-dependent transmission probabilities of particles escaping the BH’s gravitational potential~\cite{Dubinsky:2024nzo}. These GBFs modulate the otherwise blackbody Hawking spectrum and are essential for accurate predictions of observational signatures. However, computing GBFs involves solving complex differential equations for each particle species and spin in curved spacetime, which is typically a numerically intensive task. To address this, we employ SR to derive analytic approximations for the numerically computed GBFs, using the recently developed \texttt{GrayHawk} package~\cite{Calza:2025whq}. Our symbolic models not only reproduce the numerical results with high accuracy, but also offer compact, interpretable expressions suitable for fast evaluation. We demonstrate that these SR-based approximations can be used to place constraints on $f_{\rm PBH}$, the fraction of DM composed of PBHs, using measurements of the diffuse Extragalactic $\gamma$-Gay Background (EGRB). By comparing SR and numerical results, we validate the effectiveness of the former in reproducing constraints on $f_{\rm PBH}$.

This approach is particularly relevant for assessing PBHs as DM candidates. A distinctive observational signature of low-mass PBHs is their Hawking radiation, a quantum mechanical effect by which BHs lose mass through the emission of a quasi-thermal spectrum of particles~\cite{Hawking:1975vcx,Saini:2017tsz}. This radiation encompasses photons~\cite{Arbey:2019vqx,Laha:2019ssq,Ballesteros:2019exr,Cang:2021owu, Perez-Gonzalez:2023uoi}, electrons/positrons~\cite{MacGibbon:1991vc,Boudaud:2018hqb,Su:2024hrp}, neutrinos~\cite{Dasgupta:2019cae,Wang:2020uvi,Lunardini:2019zob,Calabrese:2021zfq,DeRomeri:2024zqs}, and other Standard Model particles~\cite{Stojkovic:2004hz, Carr:2009jm,Clark:2016nst,Acharya:2020jbv,Calza:2023iqa,Calza:2021czr,Calza:2023rjt} and beyond~\cite{Stojkovic:2005zh,Dong:2015yjs,Cang:2022jyc}. 
This radiation emitted by PBH is expected to produce observable effects that could be identified in cosmic and astrophysical observations~\cite{Belotsky:2014kca,Carr:2020gox, Carr:2020xqk,Auffinger:2022khh}. More precisely, the emission from PBH accumulated across cosmic history should leave a detectable signature in $\gamma$-ray surveys~\cite{Carr:2016hva, Chen:2021ngo}. Analyzing the EGRB is therefore important for assessing how much of the DM may be comprised of PBHs. Our SR-based approximation allows for efficient and accurate calculation of the radiation predicted from PBH, thereby aiding in setting constraints on $f_{\rm PBH}$.

This paper is organized as follows. Section~\ref{sec:evaporation} covers a review of Hawking radiation alongside the significance of GBFs and introduces the \texttt{GrayHawk} code employed for numerical computation. In Section~\ref{sec:pysr}, we explain the principle and methodology of \texttt{PySR}, and describe its application to the fitting of GBFs, which is referred to as \texttt{ReGrayssion} herein. Additionally, it can be accessed publicly on \texttt{GitHub} through the link: \href{https://github.com/yuanguanwen/Regrayssion}{\texttt{https://github.com/yuanguanwen/Regrayssion}}. Section \ref{sec:methodology} presents our approach to deriving constraints on $f_{\rm PBH}$ from $\gamma$-ray observations. Finally, we summarize the implications of our findings in Section~\ref{sec:conclusion}. Throughout this work, we adopt natural units $G=c=\hbar=k_B =1$.

\section{Hawking Radiation and Gray-Body Factors}\label{sec:evaporation}
In curved space-time, different observers generally disagree on what constitutes the quantum vacuum state. This arises because of the use of different time coordinates to separate positive and negative frequency modes during field quantization. While in flat space Lorentz invariance ensures all inertial observers perform such a separation consistently, in curved space-time, especially near event horizons, this uniformity is broken.  

Hawking demonstrated in 1974~\cite{Hawking:1974rv,Hawking:1974} that a stationary observer far from a BH horizon detects a nearly thermal flux of particles coming from what a freely falling observer at the horizon would describe as a vacuum state. Hawking radiation is a purely gravitational effect that causes the BH to emit particles with masses below its Hawking temperature, \(T_H \simeq M_P^2 / 8\pi M\), leading to a loss of mass and angular momentum and ultimately to BH evaporation.  

This section will outline the necessary ingredients for calculating the Hawking spectrum and the GBFs, and introduce the method we adopted.

\subsection{Hawking Spectrum}\label{sec:HawkingRadation}
A fundamental aspect of BH evaporation is the modification to the Hawking radiation spectrum due to the geometric potential.  Although the Hawking radiation emitted at the event horizon follows a black-body spectrum, the surrounding gravitational potential acts as a barrier, attenuating part of the emitted radiation. The fraction of radiation that escapes to infinity is characterized by GBFs, which are frequency- and angular-momentum-dependent transmission coefficients denoted by $\Gamma^s_{lm}(\omega)$. Here, $s$ is the field spin and $l, m$ are the angular and azimuthal numbers, respectively. These factors encode how the curved spacetime alters the radiation observed at infinity, making them essential in accurately modeling the BH emission and evaporation.

Computing GBFs requires solving a classical wave scattering problem in the BH gravitational potential. The problem is mathematically governed by quantum field perturbations in curved spacetime. Specifically, quantum field equations are generalized to a non-Minkowskian metric on which the fields' back-reaction is not considered. The use of the Newmann-Penrose (NP) formalism~\cite{Newman:1961qr} allows one to condense these equations into a single equation for the respective NP-scalars: the Teukolsky equation~\cite{1973ApJ...185..635T, Press:1973zz, Teukolsky:1974yv}. The Teukolsky equation is a second-order partial differential equation describing the propagation of perturbations of a given spin $s$ in the BH background. This paper will primarily concentrate on the Schwarzschild metric, which is typically expressed in Boyer-Lindquist coordinates as
\begin{equation}
    ds^2=-\left(1-\frac{2M}{r}\right)dt^2+\frac{dr^2}{1-\frac{2M}{r}}+r^2(d\theta^2  + \sin^2\theta d\phi^2).
\end{equation}
where $M$ is the BH mass, and the angular part is the metric of the $2$-sphere.
In this case, the Teukolsky equation takes the simple form
\begin{align}
& \left [ - \frac{r^2}{f} \partial_t^2 + s \left ( r^2 \frac{f'}{f} -2 r \right ) \partial_t \right ] \Upsilon_s \nonumber \\
&+ \left [ (s+1) (r^2 f' + 2 r f) \partial_r \right ] \Upsilon_s \nonumber \\
&+\left [ \frac{1}{\sin{\theta}} \partial_\theta (\sin{\theta}\partial_\theta) + \frac{2 i s \cot{\theta}}{\sin{\theta}} \partial_\phi \right. \nonumber \\
&\left. + \frac{1}{\sin^2{\theta}} \partial^2_\phi -s -s^2 \cot^2{\theta} \right ] \Upsilon_s \nonumber \\
&+\left [ s r^2 f'' + 4 s r f' + 2 s f \right ] \Upsilon_s=0\,.
\label{eq:teukolsky}
\end{align}
with $f(r)=1-\frac{2M}{r}$, and $' \equiv \partial_r$.

This equation admits separable solutions where the perturbation variable $\Upsilon_s$ is decomposed as follows:
\begin{eqnarray}
\Upsilon_s= \sum_{l,m} e^{-i \omega t } e^{i m \phi} S^{l}_s(\theta) R_s(r)\,,
\label{eq:upsilon}
\end{eqnarray}
here $\omega$ is the perturbation frequency, $S^s_{l,m}(\theta, \phi)=\sum S^l_s(\theta) e^{im\phi}$, satisfying the spin-weighted spherical harmonics equation ~\cite{Fackerell:1977ghw,Suffern:1983ghw,Seidel:1988ue,Berti:2005gp}. 
The radial equation reads
\begin{align}
&\frac{1}{\Delta^s}\big(\Delta^{s+1}R'_s\big)' \nonumber \\
&+\left(\frac{\omega^2r^2}{f}+2i\omega sr-\frac{is\omega r^2f'}{f}+s(\Delta''-2)-\lambda_l^s\right)R_s=0\,,
\label{eq:radialteukolsky}
\end{align}
where $\lambda_l^s\equiv l(l+1)-s(s+1)$, and $\Delta(r)\equiv r^2f(r)$.

The GBFs are computed by imposing purely ingoing boundary conditions at the horizon and extracting the transmission coefficient of the scattering problem $\Gamma^s_l(\omega)$.
The emission rate of a given particle species $i$ with spin $s$, due to Hawking evaporation, is given by~\cite{Hawking:1975iha,Page:1976df,Page:1976ki,Page:1977um}:
\begin{equation}
\frac{d^2N_i}{dtdE_i}=\frac{1}{2\pi}\sum_{l} (2l+1)\frac{n_i\Gamma^s_{l}(\omega)}{ e^{\omega/T}\pm 1} ,
\label{eq:d2ndtdei}
\end{equation}
where $n_i$ is the number of degrees of freedom of the emitted particle, $\omega = E_i$ is the mode frequency, and the GBFs $\Gamma^s_{l}(\omega)$ determine the transmission probability for each mode. The sign at the denominator accounts for the Bose-Einstein or Fermi-Dirac statistics, $-$ and $+$ respectively, while the factor $(2l+1)$ in Eq.~(\ref{eq:d2ndtdei}) arises from the degeneracy in the azimuthal number $m$ due to spherical symmetry. Eq.~(\ref{eq:d2ndtdei}) encapsulates the key role of GBFs in shaping the energy spectrum of Hawking radiation, making their accurate computation crucial for studying BH evaporation.

\subsection{Gray-Body Factors and \texttt{GrayHawk}}
Over the years, several numerical codes have been developed to calculate the Hawking radiation spectra (among other quantities related to BHs), such as \texttt{BlackHawk}~\cite{Arbey:2019mbc}, \texttt{BlackMax}~\cite{Dai:2009by}, \texttt{Charybdis}~\cite{Frost:2009cf}, and \texttt{CosmoLED}~\cite{Friedlander:2022ttk}. Among these, we adopt a modified version of \texttt{GrayHawk}~\cite{Calza:2025whq} (namely, we enlarged its capability to the $s=3/2$ field) due to its high numerical precision and systematic approach. 
\texttt{GrayHawk} solves the perturbations scattering problem using the equivalent of Eq.~(\ref{eq:radialteukolsky}) for a generic spherically symmetric metric written in Schr\"odinger-like form, and computing the GBFs across a wide range of BH parameters, as well as different perturbation field spin and energy regimes.
In the following sections, we use \texttt{GrayHawk} to generate numerical GBFs for Schwarzschild BHs, fit these results using both SR and a method which we call Human Regression (HR), described later. We then use those expressions to derive constraints on the fraction of DM composed of PBH $f_{\rm PBH}$.

\section{Symbolic Regression for GBFs}\label{sec:pysr}

Symbolic Regression (SR), a powerful machine learning technique capable of discovering concise and human interpretable mathematical relationships directly from given data~\cite{makke2024interpretable,NEURIPS2022_dbca58f3,wang2023scientific}, has recently attracted increasing interest within the machine learning community.  
In contrast to standard regression methods that use a fixed model format, SR explores a vast range of possible equations to achieve the best balance of accuracy and clarity. By defining a target variable, a collection of potential explanatory variables, and a set of available mathematical operators, SR aims to find equations that accurately reconstruct the target variable while remaining concise and understandable. These abilities enable SR to make significant advancements in theoretical research as well as in the analysis of scientific data. We use \texttt{PySR}~\cite{cranmer2023interpretablemachinelearningscience}\footnote{\href{https://github.com/MilesCranmer/SymbolicRegression.jl}{https://github.com/MilesCranmer/SymbolicRegression.jl}} to work out interpretable expressions for GBFs with applications to evaporation, which is typically complex for traditional numerical computations.

\subsection{\texttt{PySR}}\label{sec:pysr}

\texttt{PySR} is a high-performance symbolic regression package implemented in Python and Julia~\cite{Cranmer:2020wew, cranmer2023interpretablemachinelearningscience}\footnote{\href{https://github.com/MilesCranmer/PySR}{https://github.com/MilesCranmer/PySR}}. A critical element of \texttt{PySR} involves achieving a balance between precision and simplicity, in accordance with Occam's razor. Moreover, \texttt{PySR} inherently generates straightforward analytic approximations, making it an effective tool for obtaining empirical expressions in complex astrophysical scenarios.

\texttt{PySR} employs a multi-population evolutionary algorithm that systematically searches for candidate equations from a predefined set of mathematical operators and variables. 
Each symbolic expression is assessed based on a scoring mechanism that incorporates both accuracy and complexity; the  latter can be regulated by a user-defined parsimony coefficient. Accuracy is typically evaluated via a loss function, while complexity is quantified as a weighted sum of the number of variables, operators, and constants in the expression. Table~\ref{tab:operators} lists the binary operators ($+, -, \times, \div, {\rm pow}$) and unary operators (sin, cos, exp, log, sqrt, erf, erfc) adopted in this work.

A widely used loss function is the Mean Squared Error (MSE):
\begin{equation} 
\ell_{\rm pred}(E) = \sum_i w_i \times[ X_{{\rm obs}, i} - X_{{\rm pred},i}(E) ]^2, 
\end{equation} 
where $E$ is the explanatory variable, $X_{{\rm obs}, i}$ and $X_{{\rm pred},i}(E)$ denote the observed and predicted values at each data point $i$,  and $w_i$ is a weight factor, usually taken as the inverse of the observational uncertainty. This weighted loss enables the model to give more weight to higher-quality data points, improving robustness to noise and yielding more physically meaningful fits.

Traditional SR frameworks usually apply a fixed penalty to express complexity $\ell(E)$. In contrast, \texttt{PySR} allows for a dynamic, data-driven penalization scheme, which is expressed roughly as~\cite{cranmer2023interpretablemachinelearningscience}
\begin{equation} 
\ell(E) = \ell_{\rm pred}(E) \cdot \exp\left[\mathrm{frecency}(C(E))\right], \label{eq:sr_loss} 
\end{equation} 
where the `frecency' metric reflects the combined frequency and recency of expressions of complexity $C(E)$ encountered in the evolving population. This adaptive mechanism penalizes overrepresented complexities and encourages a diverse exploration of the expression space, including both simple but potentially imprecise models and complex but highly accurate ones. As a result, the search is less prone to premature convergence and better able to identify viable candidate models.
The evolutionary process of \texttt{PySR} is then guided toward interpretable models by minimizing the loss function, which balances the accuracy and complexity. 
These allow the process to prioritize high-quality data, enhancing robustness and fostering physically motivated expressions. Furthermore, \texttt{PySR} also allows for a custom formulation of the loss function tailored to the specific input data context to optimize the analytic and interpretability function. Even though \texttt{PySR} does not deterministically find a unique optimal fitting equation, it produces a collection of candidate models for different complexities, then one has to decide what complexity and expression they want. The ultimate choice is guided by scientific insight and expertise, ensuring that the derived relationship aligns with the expected physical phenomena.  

Neural networks can model complex nonlinear relationships within data. However, their lack of interpretability often constrains their application in theoretical physics~\cite{Wang:2024ixk,2025arXiv250201702M}. In contrast, deriving explicit analytic expressions that describe or approximate fundamental physical relationships is advantageous as it enhances understanding and supports further theoretical investigation. For this purpose, we use the method intended to approximate functions with an analytic expression. SR follows an evolutionary approach in which mathematical expressions are iteratively refined through a mutation-selection process. The search begins with a pool of candidate equations constructed from predefined mathematical operators. These expressions evolve over successive generations, and the most accurate and parsimonious ones survive into the next iteration. Mutations and Crossovers allow for continuous exploration of the function space.

\subsection{\texttt{ReGrayssion} }
This research utilizes \texttt{PySR}, as discussed in Sec~\ref{sec:pysr}, to identify optimal analytic approximations for GBF within Hawking radiation for the first time, which we refer to \texttt{ReGrayssion} as the pipeline we develop in this context.

\begin{table}[htbp!]
\centering
\begin{tabular}{|c|c||c|c|}
\hline
binary & complexity & unary & complexity \\
\hline
+ & 1 & sin & 3   \\
\hline
- & 1 & cos & 3   \\ 
\hline
$\times$ &2 & exp & 3 \\
\hline
$\div$ & 2 & log & 3 \\
\hline
pow & 3 & erf/erfc &4 \\
\hline
\end{tabular}
\caption{The operators and complexities are used to construct the symbolic expressions in \texttt{ReGrayssion}.}
\label{tab:operators}
\end{table}

Coming up with analytical formulas for GBFs is a challenge. Many studies have proposed a solution to this problem based on the WKB approximation ~\cite{Iyer:1986np, Konoplya:2019hlu, Dubinsky:2024nzo}. However, this method is valid in the large $l$ limit, although percent level precision in the determination of GBFs can be accomplished even for low $l$ by introducing higher order corrections. Nevertheless, it has been shown that even these higher-order GBFs are still not accurate enough to properly reproduce the Hawking evaporation spectrum \cite{Konoplya:2023moy,Pedrotti:2025upg}. Other approaches have been proposed, based on the solution to the connection problem of the confluent Heun equation in terms of the explicit expression of irregular Virasoro conformal blocks as sums over partitions via the Alday, Gaiotto, and Tachikawa correspondence, see for example ~\cite{Bonelli:2021uvf,CarneirodaCunha:2015hzd,CarneirodaCunha:2015qln,CarneirodaCunha:2019tia,BarraganAmado:2021uyw,Dubinsky:2024nzo}. These calculations, although exact, result in very complex equations, which largely depend on the black hole model considered.  

For these reasons, we implement here the \texttt{ReGrayssion} pipeline to mining analytical but concise approximations to the Schwarzschild GBFs. We then assess their accuracy by comparing the derived formulae with the numerically computed GBFs from \texttt{GrayHawk}.

\subsection{Methodology: GBF  analytic approximations} \label{analyticGBF}

We perform these analytic approximations with a hybrid approach, combining \texttt{ReGrayssion} to discover general trends and template fitting for precision - which allows us to derive both data-driven and physics-informed expressions for the GBF. By implementing SR alongside the human regression function, we aim to develop precise, interpretable, and computationally efficient approximations for application in BH physics.  Hence, the GBFs are then fitted using two different analytical approximations:   

\begin{itemize}
\item \textbf{Symbolic Regression (SR)}: Instead of assuming an explicit functional form, SR employs genetic algorithms to directly discover analytical expressions from numerical data. This method balances accuracy and simplicity, potentially uncovering underlying structures that are not immediately apparent using conventional fitting techniques.

\item \textbf{Human Regression (HR)}: Inspired by the expression found with SR we propose a similar function whose parameters are optimized using the \texttt{SciPy-Optimize}e~\footnote{\href{https://docs.scipy.org/doc/scipy/reference/optimize.html}{https://docs.scipy.org/doc/scipy/reference/optimize.html}} package.
\end{itemize}

During the analytic fitting, we consider modes up to $l=4$ for bosons and $l=7/2$ for fermions, numerically computed with  \texttt{GrayHawk}. Higher modes do not significantly contribute to the final spectrum~\cite{Calza:2022ljw,Calza:2023gws, Calza:2024fzo}. To prevent overfitting in SR, we introduce Gaussian noise in $\Gamma^s_{\omega}$, following a normal distribution $\mathcal{N}(\mu, \sigma^2)$, where. 

\begin{align}\label{eq:error}
\tilde{\omega}_i &= \omega_i + \mathcal{N}(0, \epsilon*\omega_i), \\
\tilde{\Gamma}_i &= \Gamma_i + \mathcal{N}(0, \epsilon*\Gamma_i).
\end{align}

\begin{figure}[htbp]
\centering
\includegraphics[width=\linewidth]{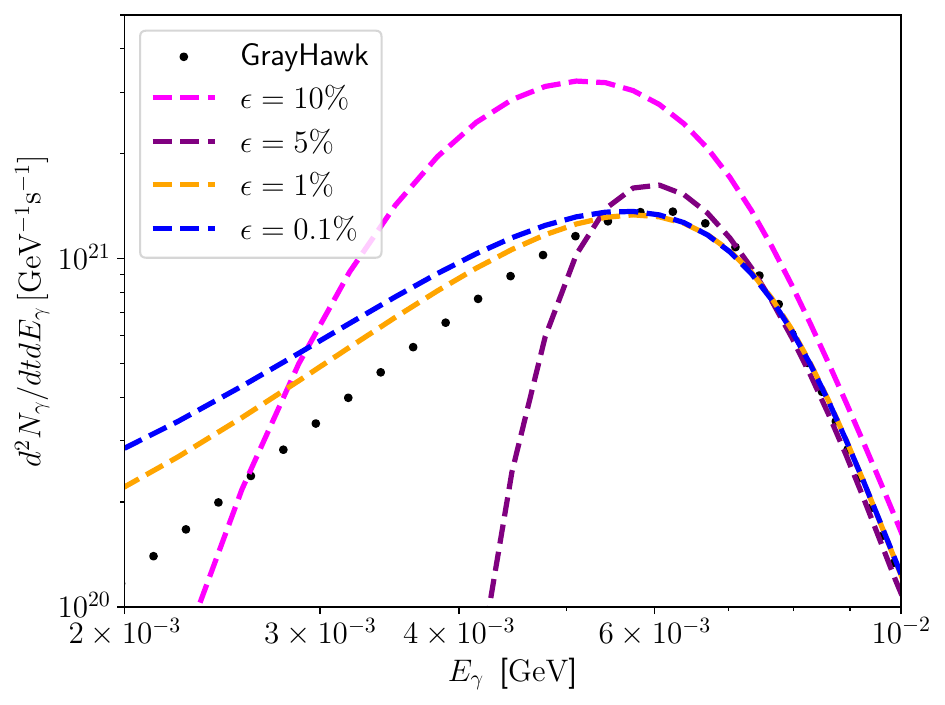}
\caption{Photon spectra from the evaporation of a Schwarzschild PBH with mass $M_{\rm PBH} = 10^{16}~\mathrm{g}$.
The spectra are computed using SR models trained on GBFs $\Gamma(\omega, \ell)$ under different noise levels.}
\label{fig:sr_errors}
\end{figure}

We conducted an investigation of different noise levels, specifically $\epsilon = 10\%, 5\%, 1\%$ and $0.1\%$.  Fig.~\ref{fig:sr_errors} showcases the photon spectra derived from the analytic functions of GBFs, which were fitted across varying noise levels.
The results indicate that as noise levels increase, it becomes increasingly difficult for \texttt{ReGrayssion} to discern suitable analytic expressions. In particular, when noise reaches $\epsilon \geq 10\%$, the chaotic behavior of the data significantly prevents SR from formulating reliable expressions. Detailed results are extensively illustrated in Fig.~\ref{fig:sr_4errs} found in the Appendix. 
After extensive tests, we find that \texttt{ReGrayssion} produces stable and consistent results when the noise level is $\epsilon \sim 1\%$ and decreasing the noise to 0.1\% results in a worsening of the spectra. Based on these results, we adopt the noise level $\epsilon = 1\%$ for the GBF analytic approximations. When we fit the numerical data with fixed $s$, \texttt{ReGrayssion} recommended the following SR expression for GBFs
\begin{equation}
\Gamma^{s}_{l} (\omega)_{\rm SR} = {\rm erf} [ {\rm erfc}(a+b*l+c*\omega) ], 
\label{eq:sr}
\end{equation}
where `erf' and `erfc' denote the error function and its complementary function, respectively. The parameters associated with each variable, energy $\omega$ and angular number $l$, namely $c$ and $b$, as well as $a$, which collectively modify the behavior of the error function to accurately approximate $\Gamma^{s}_{l} (\omega)$. The recommended values for $a$, $b$, and $c$ are provided by \texttt{ReGrayssion}.
Taking inspiration from the concise expression recommended by \texttt{ReGrayssion}, we propose an alternative more concise expression, whose complexity is 19 similar to that in Eq.~(\ref{eq:sr}), but offers a more concise HR expression:  
\begin{equation}
\Gamma^{s}_{l} (\omega)_{\rm HR} =\frac{1}{2} [ {\rm erf}(\alpha+\beta*l+\gamma*\omega) +1 ],
\label{eq:tf}
\end{equation}
Table~\ref{tab:manualcoeff} presents the optimal coefficients for Eq.~(\ref{eq:sr}) and Eq.~(\ref{eq:tf}) across various spin states: $s=0, 1/2, 1, 3/2, 2$. 

\begin{table}[htbp]
\centering
\begin{tabular}{|c||c|c|c||c|c|c|}
\hline
 & \multicolumn{3}{c||}{$\Gamma (\omega)_{\text{SR}}$} & \multicolumn{3}{c|}{$\Gamma (\omega)_{\text{HR}}$}  \\
\hline
$s$ & $a$ & $b $ & $b $ & $\alpha$ & $\beta $ & $\gamma $ \\ 
\hline
0   & 1.688  & 1.880  & -10.057  & -1.586  & -2.530  & 13.533  \\ 
\hline
1/2 & 1.461  & 1.960  & -10.149  & -1.279  & -2.636  & 13.651  \\
\hline
1   & 1.097  & 2.059  & -10.293  & -0.790  & -2.770  & 13.847  \\
\hline
3/2 & 0.535  & 2.200  & -10.454  & -0.034  & -2.960  & 14.064  \\
\hline
2   & -0.049 & 2.278  & -10.510  & 0.752   & -3.065  & 14.140  \\ 
\hline 
\end{tabular}
\caption{Coefficients of the GBFs for different spin values $s$, obtained using the SR in E .~(\ref{eq:sr}) and HR in Eq.~(\ref{eq:tf}). We use the $\epsilon=1\%$ noise level during the fitting processes. }
\label{tab:manualcoeff}
\end{table}

By default, variable and parameter complexities are set to 1, making the complexity of the polynomial ($a+b*l+c*\omega$) 10. Thus, the total complexity of Eq.~(\ref{eq:sr}) becomes 18 with the addition of `erf' and `erfc'(see Fig.~\ref{fig:expressiontree}). Although we remove the `erfc' and add two operators ($\times, +$) and parameters, increasing the complexity to 19, Eq.~(\ref{eq:tf}) remains more interpretable than Eq.~(\ref{eq:sr}).

It is important to recognize that the analytic expressions generated by \texttt{ReGrayssion} are diverse, with some being closer to the data due to added complexity (`overfitting'), while others are too simple to miss some main features in the data (`underfitting'). The \texttt{ReGrayssion} scoring system ranks these expressions by balancing their loss function and complexity, surpassing traditional methods like Akaike and Bayesian information criteria~\cite{Akaike:1974vps, Thrane:2018qnx} in efficiency, which are widely used in comparing different complexity models.  We also explored similar structures with higher order polynomials in Sec.~\ref{sec:observation}, which are also inspired by Eq.~(\ref{eq:sr}).

\begin{figure*}[htbp]
\centering
\includegraphics[width=0.8\linewidth]{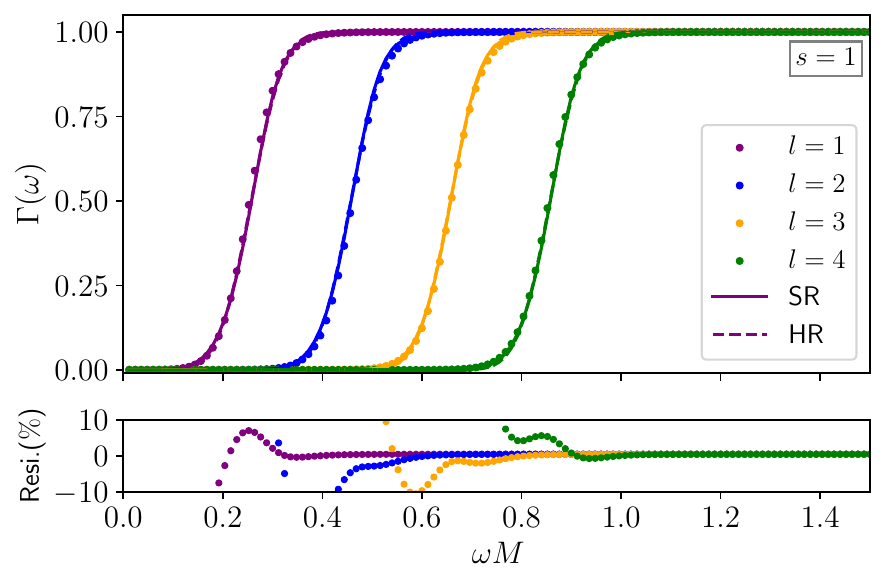}
\caption{GBFs for Schwarzschild BHs with $s=1$. Different colors points indicate different multipole numbers $l$ as computed by \texttt{GrayHawk}~\cite{Calza:2025whq}. Solid and dashed curves represent analytic fits from SR in Eq.~(\ref{eq:sr}) and HR in Eq.~(\ref{eq:tf}), respectively. Most of the residuals with numerical data of the SR are within 10\%.}
\label{fig:GBFs=1}
\end{figure*}

Fig.~\ref{fig:GBFs=1} illustrates the GBFs for Schwarzschild BHs with $s=1$. Each color point indicates the numerical results from \texttt{GrayHawk} for different multipole numbers $l$, whereas the solid and dashed lines (which on the plot superimpose resulting in a single line) represent the analytical approximation from SR (Eq.~\ref{eq:sr}) and HR (Eq.~\ref{eq:tf}), respectively. The residuals $(\Gamma(\omega)_{\rm GrayHawk} - \Gamma(\omega)_{\rm SR})/\Gamma(\omega)_{\rm GrayHawk}$ quantify the deviation of the fits from the numerical results, demonstrating the reliability of the derived expressions. By systematically comparing these fitting methods, we highlight how both SR and HR provide reliable expressions for the GBFs.

In short, our \texttt{ReGrayssion} pipeline employs a genetic algorithm to search for analytical expressions, striking a balance between accuracy and simplicity by optimizing both MSE loss and the complexity score. In the HR stage, we start from the SR motivated functional forms, and subsequently fit the parameters through least squares fitting. This systematic approach constructs analytical approximations for the GBFs.

\section{Constraining PBH as DM Using  Analytical GBFs}\label{sec:methodology}

The possibility of PBH accounting for the entire DM budget is highly limited by numerous observations~\cite{Katz:2018zrn,Bai:2018bej,Smyth:2019whb,Coogan:2020tuf,Ray:2021mxu,Auffinger:2022dic,Ghosh:2022okj,Miller:2021knj,Branco:2023frw,Bertrand:2023zkl,Tran:2023jci,Gorton:2024cdm,Dent:2024yje,Tamta:2024pow,Tinyakov:2024mcy,Loeb:2024tcc}. The only remaining parameter space where PBHs might constitute all DM is the asteroid mass window, approximately spanning PBH masses from $10^{17} {\rm g}$ to $10^{23} {\rm g}$~\cite{Montero-Camacho:2019jte,Coogan:2020tuf,Auffinger:2022khh, Dent:2024yje,Dai:2024guo}. PBHs with lower masses would either have vanished due to Hawking evaporation or emitted too many MeV $\gamma$-rays to be consistent with the observations.

Following the discussion in Sec.~\ref{sec:HawkingRadation}, we take the analytic expressions of GBFs for photons ($s=1$) in the Schwarzschild case as an example to place constraints on $f_{\rm PBH}$. The results obtained test the validity of our approximation.

\subsection{Photon Spectrum from PBH Evaporation }
We focus exclusively on the Hawking primary photons, neglecting secondary contributions from unstable particle decays~\cite{Calza:2024fzo, Calza:2024xdh}.  Angular modes up to $l=4$ are considered, with higher order modes found to have a negligible impact on the final spectrum. We adopt the following assumptions: 
\begin{itemize}
    \item  PBHs are isotropically distributed on sufficiently large scales;
    \item  the PBH population follows a monochromatic mass distribution where all PBHs share the same mass $M_{\rm PBH}$;
    \item  PBHs cluster in the galactic halo similarly to other DM components;
    \item  the primary photon spectrum dominates in the integral. 
\end{itemize}
     
The present-day primary photon flux $I(E_{\gamma 0})$ from PBH evaporation is given by the integrated sum over all PBHs in the Universe that have been emitting photons since recombination. The integral,  taking into account the scaling of energy and number density with redshift~ is given by \cite{Carr:2009jm}:
\begin{equation}
I(E_{\gamma 0}) = A_I \int_0^{z_{\star}} \frac{dz}{H(z)} \frac{d^2 N_{\gamma}}{dt dE_{\gamma}}(M_{\rm PBH}, (1+z)E_{\gamma 0}),
\label{eq:IE0}
\end{equation}
where ${d^2 N_{\gamma}}/{dt dE_{\gamma}}$ is given in Eq.~(\ref{eq:d2ndtdei}), and the normalization factor $A_I$ is
\begin{equation}
A_I = \frac{c}{4\pi} n_{\rm PBH}(t_0)E_{\gamma 0}.
\end{equation}
Here $H(z)$ is the Hubble expansion rate, $z_{\star}$ is the redshift of recombination, and $E_{\gamma 0}$ denotes the photon energy observed today. In Eq.~(\ref{eq:IE0}) we assumed that throughout the integration $M_{\text{PBH}}$ remains constant, despite the evaporation process responsible for the photons' emission \cite{Carr:2016hva,Carr:2016drx,Carr:2020xqk}. 
The current PBH number density $n_{\rm PBH}(t_0)$ is constrained by matching the theoretical un-normalized flux $I(E_{\gamma 0})/n_{\rm PBH}(t_0)$ to measurements of the EGRB, such as HEAO-1~\cite{Gruber:1999yr}, COMPTEL~\cite{Schoenfelder:2000bu}, and EGRET~\cite{Strong:2004ry}. Specifically, to constrain the abundance of PBHs across the mass window $10^{15}{\rm g} \lesssim M_{\text{PBH}} \lesssim 10^{17}{\rm g}$, we compute $I(E_{\gamma 0})$ for different values of $M_{\text{PBH}}$, assuming a monochromatic distribution and adjusting $n_{\rm PBH}(t_0)$ to maintain consistency with the EGRB data. Finally, the PBH fraction of DM, $f_{\rm PBH}$, is constrained by ~\cite{Calza:2024fzo, Calza:2024xdh, Calza:2025whq, Arbey:2025dnc}
\begin{equation}
f_{\rm PBH} (M) \equiv \frac{\Omega_{\rm PBH}}{\Omega_{\rm DM}} = \frac{n_{\rm PBH}(t_0) M_{\rm PBH}}{\rho_{\rm crit,0}\Omega_{\rm DM}},
\label{eq:fpbh}
\end{equation}
where $\rho_{\rm crit,0}=3H_0^2/8\pi G$ is the critical density. We assume a $\Lambda$CDM cosmology with cosmological parameters from the Planck experiment~\cite{Planck:2018vyg}. We notice that the findings remain consistent with reasonable variations in these parameters.
This framework provides a precise evaluation of PBH abundance constraints from Hawking evaporation and EGRB observations, further discussed in Sec.~\ref{sec:observation}.

\subsection{Observational Constraints on $f_{\rm PBH}$ }\label{sec:observation}
\begin{figure}[htbp]
\centering
\includegraphics[width=0.98\linewidth]{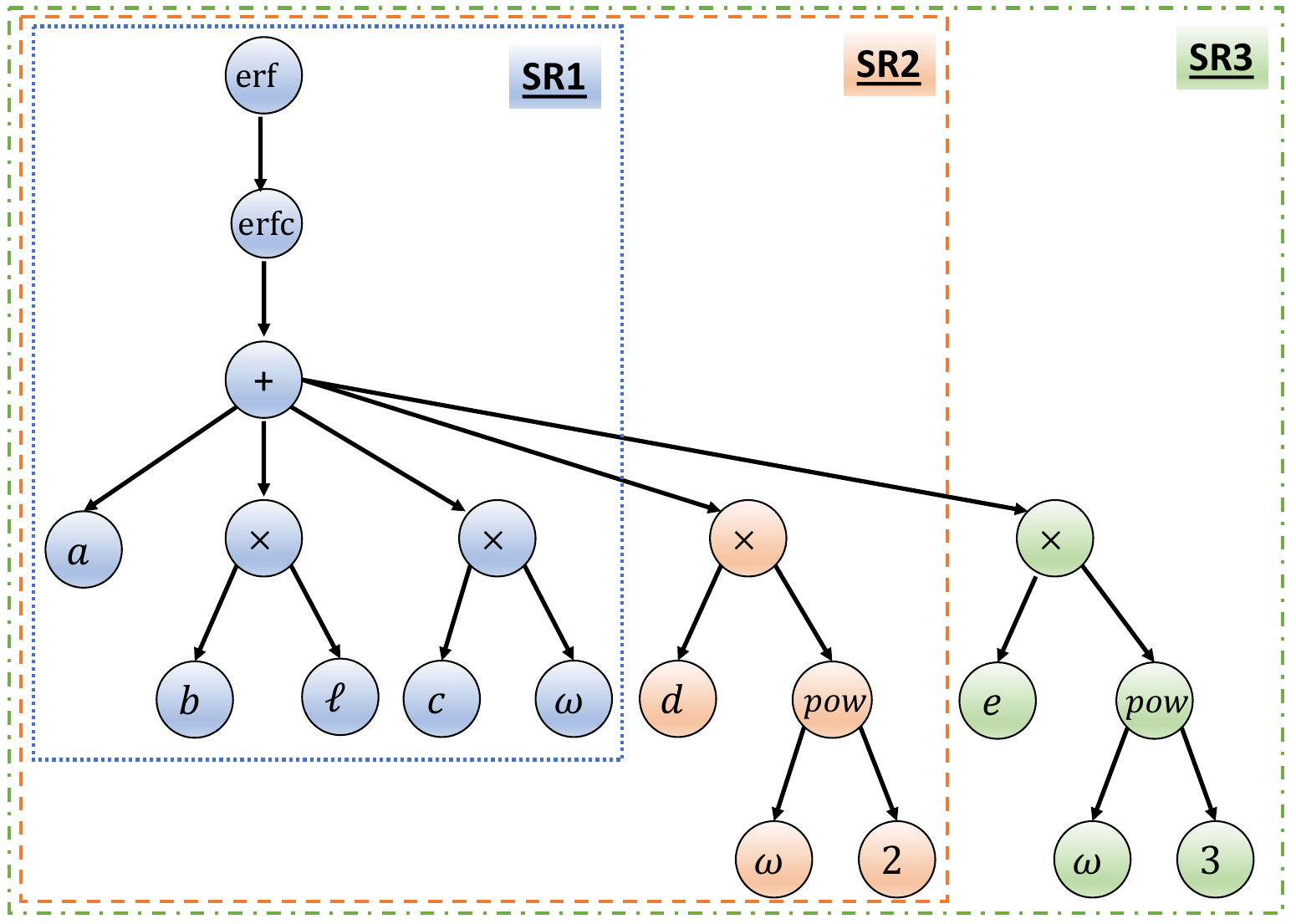}
\caption{Expression trees corresponding to the SR models in Eq.~(\ref{eq:sr123}). The default complexity of variable and constant/parameter is 1, and the operators' complexities are listed in Table~\ref{tab:operators}. Total complexities are 18 (SR1), 26 (SR2) and 34 (SR3).}
\label{fig:expressiontree}
\end{figure}

Here we focus only on the $s=1$ case and use the analytic GBF expressions previously derived in (\ref{eq:tf}), which, from now on, we will denote by SR1, to obtain the Hawking photon spectrum. In addition, we introduce two higher-order extensions of SR1. Namely, SR2 and SR3, and use them for the same purpose. Fig.~\ref{fig:expressiontree} displays the expression trees for SR1, SR2, and SR3, with complexities of 18, 26, and 34, respectively. 
The functional forms of SR1, SR2 and SR3 read  
\begin{equation}
\Gamma^{s=1}_{l} (\omega) =
\begin{cases}
    {\rm erf} [{\rm erfc} (a + b l +c\omega)] \quad  {\rm SR1}, \\
    {\rm erf} [{\rm erfc} (a + b l +c\omega + d\omega^2)] \quad {\rm SR2},\\
    {\rm erf} [{\rm erfc} (a + bl +c\omega + d\omega^2 + e\omega^3)]\quad {\rm SR3}. \\
\end{cases}
\label{eq:sr123}
\end{equation}
where the model coefficients are listed in Table~\ref{tab:photoncase} .

\begin{table}[htbp]
\centering
\renewcommand{\arraystretch}{1.2}
\begin{tabular}{|c|c|c|c|c|c|}
\hline
Model & $a$ & $b$ & $c$ & $d$ & $e$ \\
\hline\hline
SR1 & 1.097 & 2.059 & -10.293 & -  & -    \\
\hline
SR2 & 0.957 & 2.065 & -9.707  & 0.554 & - \\
\hline
SR3 & 0.928 & 2.068 & -9.523 & -0.956 & 0.242  \\
\hline
\end{tabular}
\caption{GBF coefficients for photon emission, obtained through different SR models in Eq.~(\ref{eq:sr123}).}
\label{tab:photoncase}
\end{table}

We compare these results with the numerical ones from \texttt{GrayHawk} \cite{Calza:2025whq}, concluding that, although the inclusion of higher-order terms increases complexity, in SR2 and SR3, the additional coefficients do not significantly affect stability or precision.

\begin{figure}[htbp]
\centering
\includegraphics[width=\linewidth]{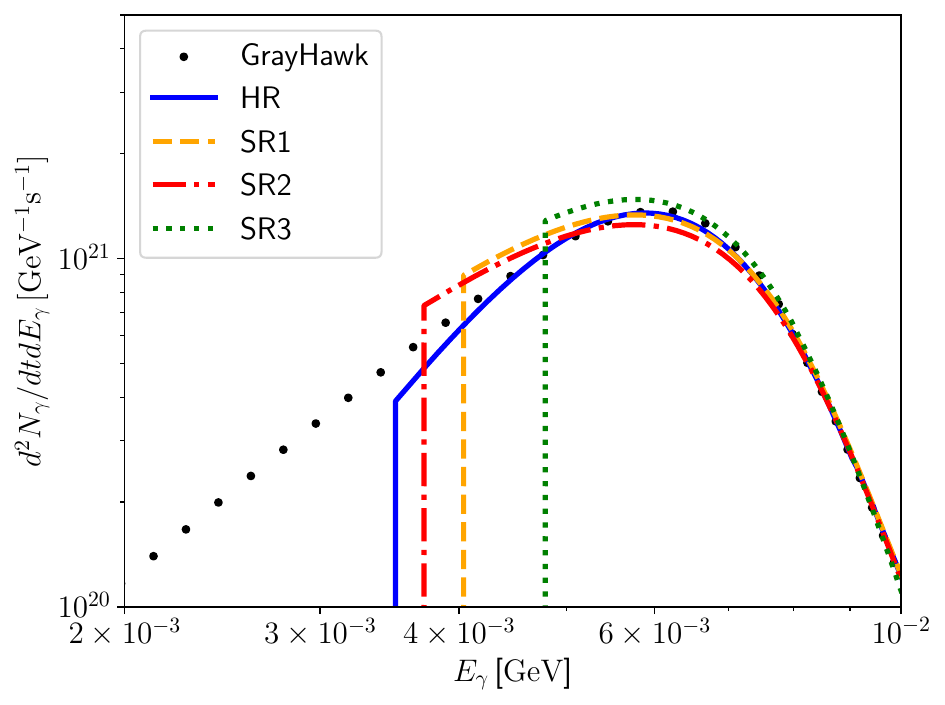}
\caption{Photon spectra from Schwarzschild PBH evaporation with $M_{\rm PBH} = 10^{16} {\rm g}$. Black points show \texttt{GrayHawk} numerical results, the solid line is the HR fit Eq.~(\ref{eq:tf}), while the dashed orange, dashed-dotted red, and dotted green lines correspond to the SR1, SR2 and SR3. }
\label{fig:spectra}
\end{figure}

Fig.~\ref{fig:spectra} presents the primary photon spectrum of a Schwarzschild PBH with mass $M_{\rm PBH} = 10^{16}{\rm g}$ calculated from the GBFs obtained through the different approaches. The black dots are the numerical results from \texttt{GrayHawk}, while the solid blue, yellow dashed, red dashed-dotted, and green dotted curves come from HR, SR1, SR2, and SR3 fits, respectively. All models reproduce the peak features well, although they show deviations in the low-energy regime, which can impact constraints derived from soft photon observations. To account for this, we have used a Heaviside step function $\Theta (\omega - \omega_{25\%})$ to cut the GBFs, setting them to 0 once the residuals exceed 25\%.

\begin{figure}
\centering
\includegraphics[width=0.95\linewidth]{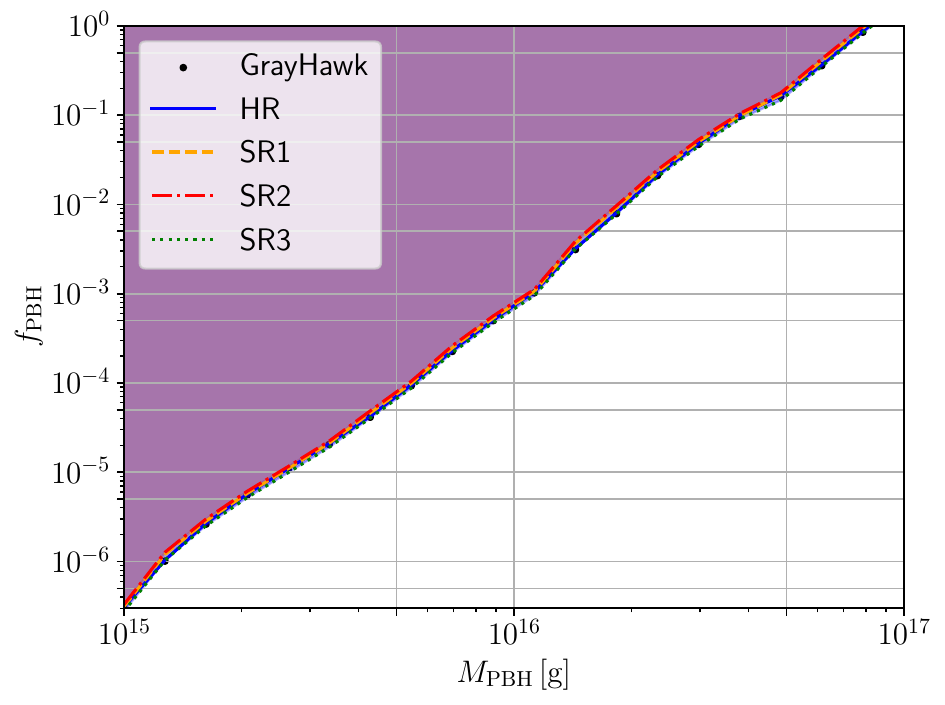}
\caption{Upper-limits on $f_{\rm PBH}$ as a function of PBH mass $M_{\rm PBH}=10^{16}$g. Same as FIG~\ref{fig:spectra}, the solid line corresponds to the HR fit, while the dashed, dashed-dotted and dotted lines represent the SR1, SR2, and SR3 model from Eq.~(\ref{eq:sr123}). The consistency of the constraints despite spectra deviations suggests that the SR/HR models capture the dominant features necessary for setting robust limits. }
\label{fig:fpbh}
\end{figure}

We derive our constraints on $f_{\rm PBH}$ making use of the observational data from HEAO-1~\cite{Gruber:1999yr}, COMPTEL~\cite{Schoenfelder:2000bu}, and EGRET~\cite{Strong:2004ry}. Despite being dated, these datasets offer the most reliable measurements across the relevant energy band. Alternative probes such as the Galactic gamma-ray background or positron fluxes suffer from greater astrophysical uncertainties and model dependencies, making the ERGB constraints more robust ones.
Fig.~\ref{fig:fpbh} shows the resulting upper limits on $f_{\rm PBH}$ as a function of PBH mass. The SR-based fits yield constraints consistent with those from the HR model and the numerical one, confirming that SR can effectively reproduce the key spectral features that drive these bounds. While deviations exist in $\Gamma (\omega,l)$, particularly at low energies, they do not significantly affect the derived limit.

We have investigated the constraints on $f_{\rm PBH}$ using both the HR and SR-based approximations for GBFs. The consistency of the limits across all fits in Fig.~\ref{fig:fpbh} confirms that these analytic approximations successfully capture the features that are relevant in setting the constraints.

\section{Conclusion}\label{sec:conclusion}
The era of ``big data" in astronomy and cosmology has ushered in a wealth of high-quality observational datasets, demanding novel analyses and interpretation techniques. Symbolic Regression (SR), a growing branch of machine learning, offers a promising approach to uncover analytic expressions in data.
Historically, scientists have engaged in an informal version of this process--deriving empirical laws through intuition and experimentation, as exemplified by Kepler, Planck, and others. Modern SR automates this paradigm, enabling the systematic exploration of vast functional spaces far beyond the reach of human intuition.

In this work, we developed the \texttt{ReGrayssion} pipeline, is a \texttt{PySR} based pipeline, to work out analytical approximations for Gray-Body Factors (GBFs) relevant to Black Hole (BH) evaporation in the Schwarzschild scenario. These symbolic expressions significantly reduce computational costs while being sufficiently accurate for the purposes of observational constraints.

Using our SR- and HR-derived GBF, we computed the Hawking radiation spectrum and derived constraints on the fraction of Dark Matter (DM) composed of Primordial Black Holes (PBHs) $f_{\rm PBH}$, based on extragalactic $\gamma$-ray background measurements. The consistency of the constraints obtained from both methods demonstrates that this approach can capture the main features relevant for the observational constraints.

In summary, we have developed compact, human-interpretable expressions for Schwarzschild GBFs using SR-based approaches. The SR-generated formulas strike a favorable balance between speed and accuracy for the specific purposes of studying PBH. Future research aims at extending this framework to Kerr BH and to refining methods for an improved treatment of energy-dependent spectral characteristics.

\begin{acknowledgments}
\noindent We are especially grateful to Sunny Vagnozzi for fruitful discussion and help during this work, and we also thank Miles Cranmer for the public \texttt{PySR} package. We acknowledge support from the Istituto Nazionale di Fisica Nucleare (INFN) through the Commissione Scientifica Nazionale 4 (CSN4) Iniziativa Specifica ``Quantum Fields in Gravity, Cosmology and Black Holes'' (FLAG). G.Y. and M.C. acknowledge support from the University of Trento and the Provincia Autonoma di Trento (PAT, Autonomous Province of Trento) through the UniTrento Internal Call for Research 2023 grant ``Searching for Dark Energy off the beaten track'' (DARKTRACK, grant agreement no.\ E63C22000500003, PI: Sunny Vagnozzi).
\end{acknowledgments}

\bibliography{reference}

\begin{appendix}

\section{Fitting GBFs with Symbolic Regression}
\begin{figure*}[htbp]
\centering
\includegraphics[width=0.48\textwidth]{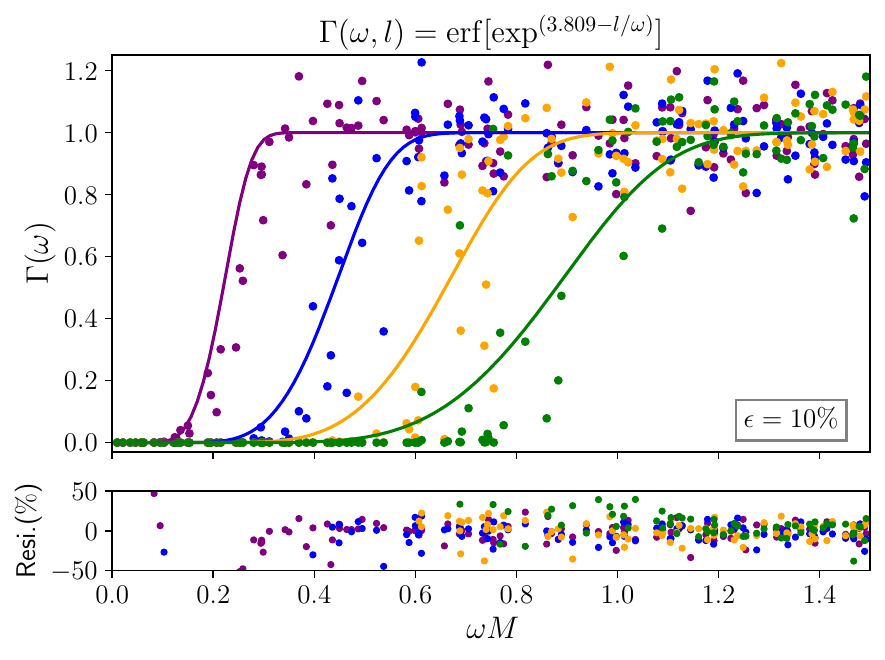} 
\includegraphics[width=0.48\textwidth]{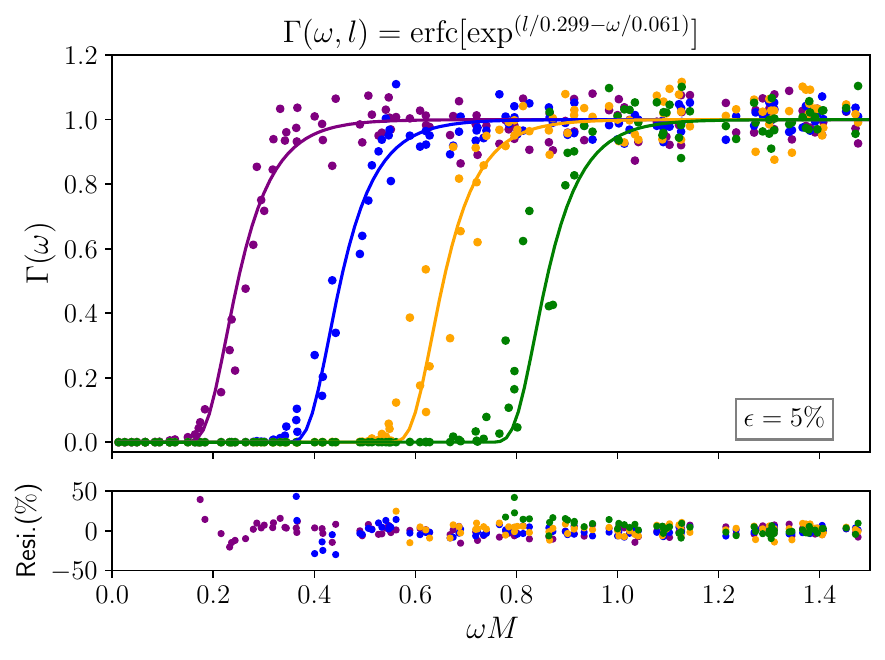}
\includegraphics[width=0.48\textwidth]{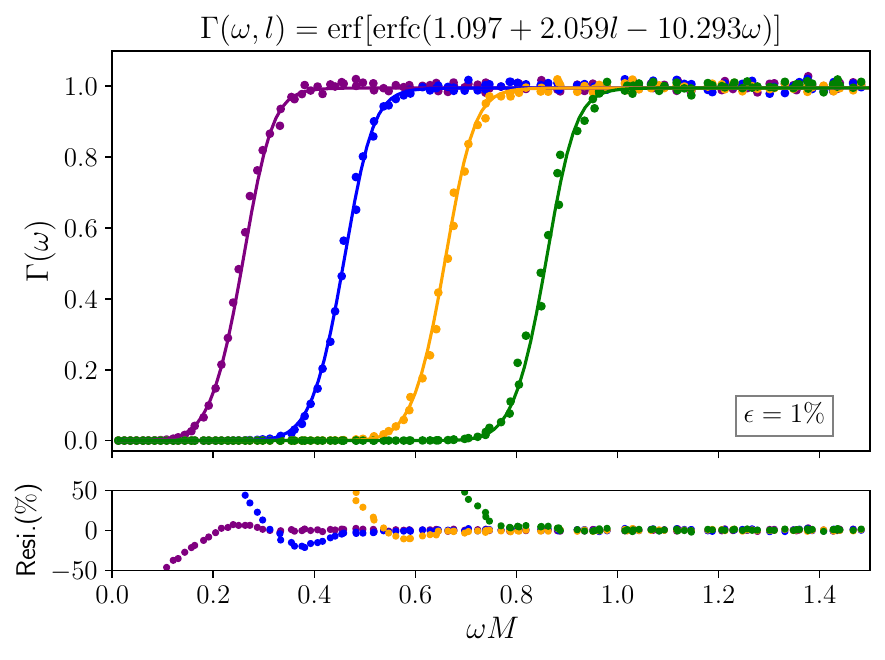} 
\includegraphics[width=0.48\textwidth]{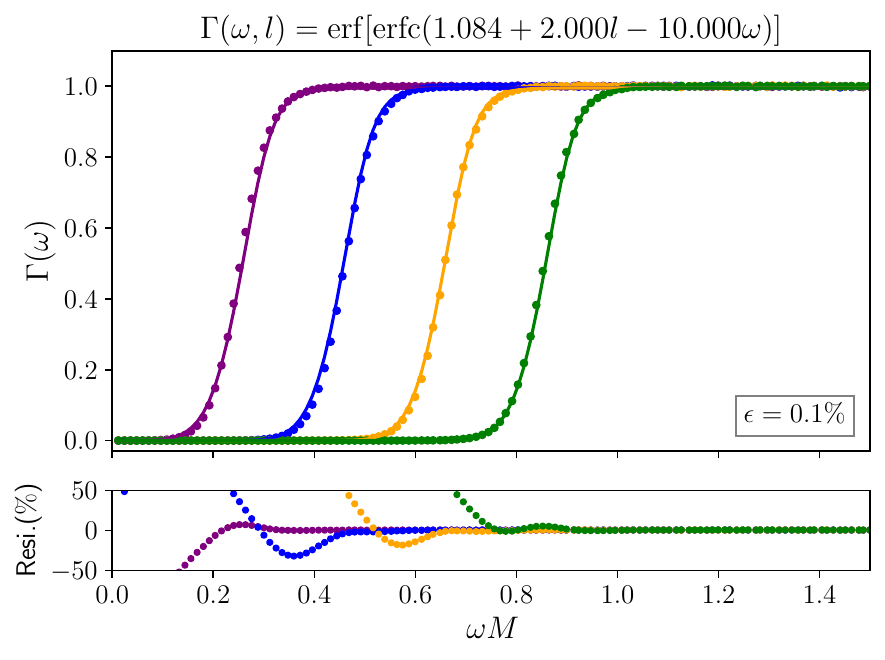}
\caption{The four subplots depicting various fits of $\Gamma (\omega, l)$ using symbolic regression (SR), each subjected to different input error rates: $\epsilon = 10\%, 5\%, 1\%$, and $0.1\% $ as described in Eq.~(\ref{eq:error}). Data points embedded with errors are plotted, alongside lines generated by functions indicated in the subplots’ titles. The colors purple, blue, orange and green correspond to cases where $l=1, 2, 3, 4$, respectively. The figure illustrates that as the magnitude of errors increases, it becomes increasingly challenging for the SR technique to derive the analytic expressions. }
\label{fig:sr_4errs}
\end{figure*}

\begin{figure*}[htbp!] 
\centering 
\includegraphics[width=0.48\textwidth]{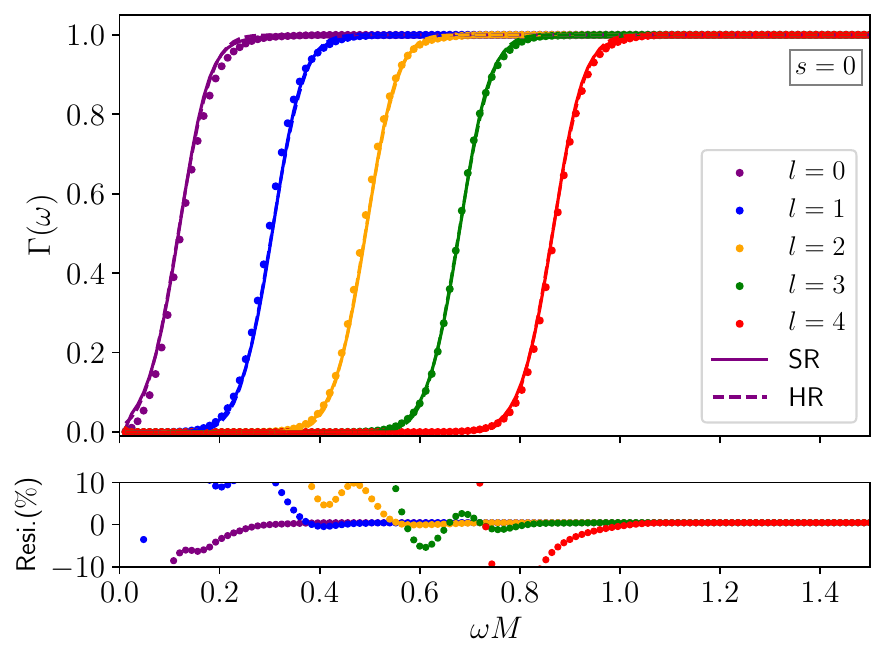} 
\includegraphics[width=0.48\textwidth]{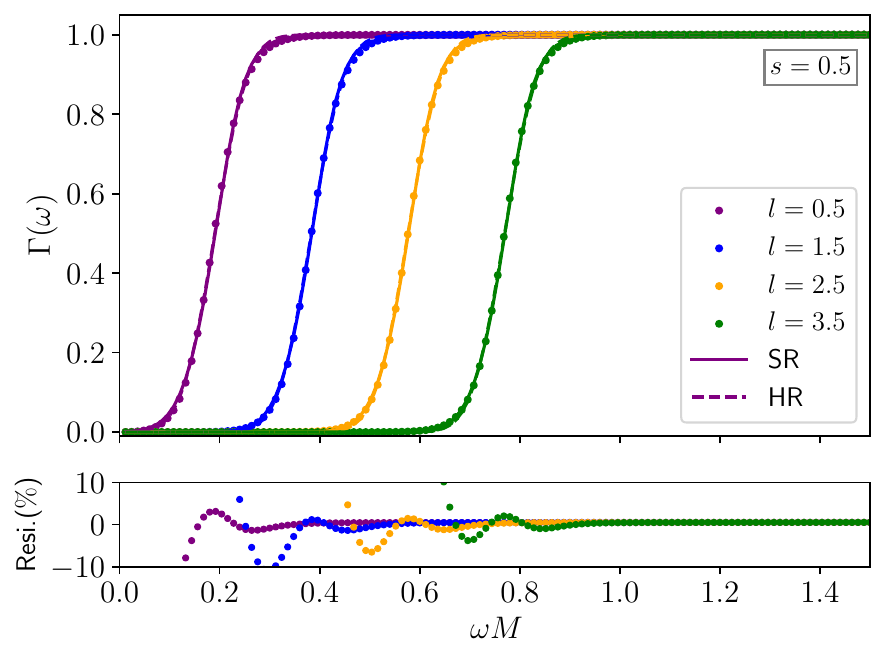}
\includegraphics[width=0.48\textwidth]{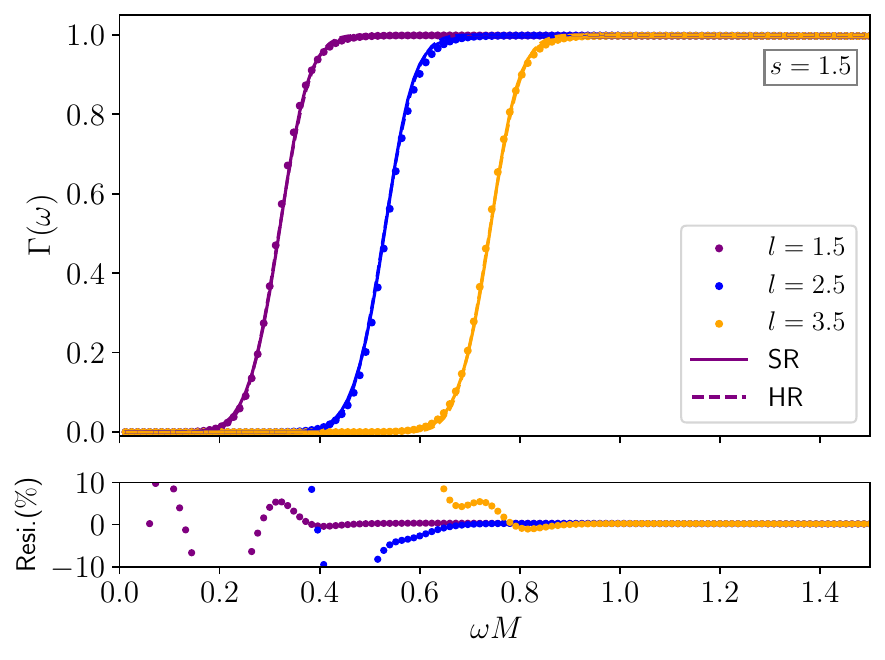} 
\includegraphics[width=0.48\textwidth]{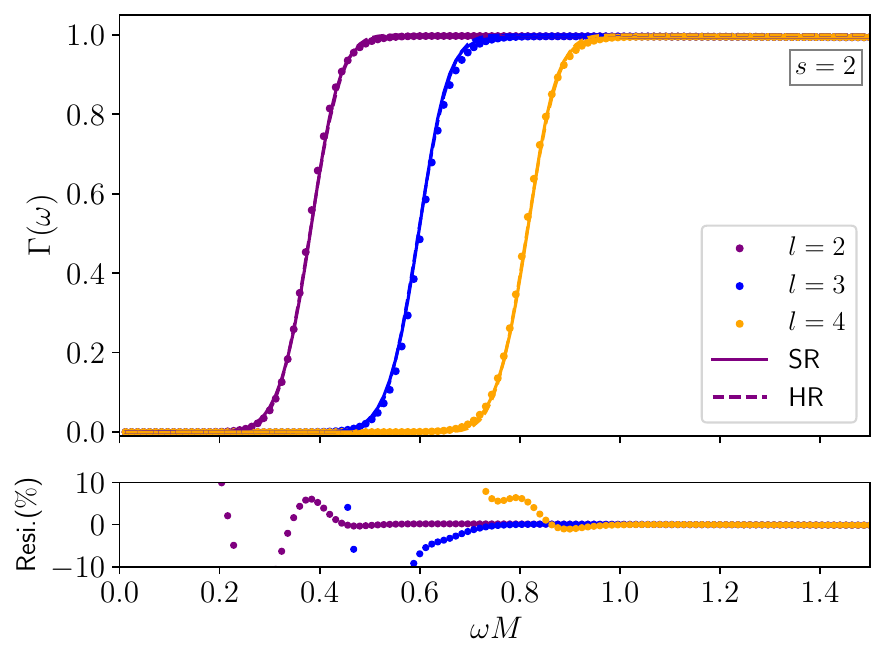}
\caption{The GBFs for Schwarzschild black holes are evaluated for spins $s=0, 1/2, 3/2, 2$. In this visualization, each different color of points corresponds to a separate multipole number $l$, derived from numerical calculations using \texttt{GrayHawk}~\cite{Calza:2025whq}. In contrast, the solid and dashed lines illustrate the analytic fits derived from SR and HR as expressed in Eq.~(\ref{eq:tf}) and Eq.~(\ref{eq:sr}), respectively. Notably, most of the SR discrepancies in the residuals is limited to only 10\%, highlighting its strong concordance with the empirical data.} 
\label{fig:Gamma_s}
\end{figure*}

Fig~\ref{fig:sr_4errs} shows various fits of $\Gamma (\omega, l)$ obtained using SR, and subjected to different error rates, namely $\epsilon = 10\%, 5\%, 1\%$, and $0.1\% $ as described in Eq.~(\ref{eq:error}). The points denote numerical data with errors, and the lines depict fitted functions. Blue, orange, green, and red correspond to $l=1, 2, 3, 4$, respectively. As noise increases, the identification of analytic expressions by SR decreases. For $\epsilon \geq 10\%$, more noisy data hinder the reliable derivation of the SR formula. Fig.~\ref{fig:Gamma_s} illustrates GBFs fitting results for Schwarzschild BH evaporating particles with spins $s=0, 1/2, 3/2, 2$. Points are \texttt{GrayHawk} numerical data, while solid and dashed lines show analytical fits of SR and HR. Residuals indicate fit deviations, which confirm the reliability of the expression.

\end{appendix}

\end{document}